%% file: hadron2011.tex
\documentclass[a4paper,11pt]{article}

\usepackage{contribution}

\input{econfmacros}
\input{contributionmacros}

\begin{document}

\input{contribution}

\end{document}

%% file: econfmacros.tex
\newcommand{\weblink}[2][]{%
    \ifthenelse{\equal{#1}{}}%
    {\textnormal{\url{#2}}}%
    {\textnormal{\href{#2}{#1}}}%
}

\newcommand{\acknowledgements}[1]{%
  \bigskip\bigskip
  \textsf{\textbf{\Large Acknowledgements}} \\[2ex]
  {#1}
  \bigskip
}

\def\beq{\begin{equation}}
\def\eeq#1{\label{#1}\end{equation}}
\def\eeqn{\end{equation}}

\def\beqa{\begin{eqnarray}}
\def\eeqa#1{\label{#1}\end{eqnarray}}
\def\eeqan{\end{eqnarray}}

\let\bar=\overbar

\def\Dslash{\not{\hbox{\kern-4pt $D$}}}
\def\dslash{\not{\hbox{\kern-2pt $\del$}}}

\def\msb{{\bar{\ssstyle M \kern -1pt S}}}

%% file: contributionmacros.tex
\newcommand{\contribution}[7][]{%
  \clearpage
  \thispagestyle{plain}
  \ifthenelse{\equal{#1}{}}
  {\hypersetup{pdftitle={#2}}}
  {\hypersetup{pdftitle={#1}}}
  \hypersetup{pdfauthor={{#3} {#4}}}
  {\centering\normalfont\LARGE\bfseries\sffamily #2 \par\nobreak}
  \lhead{}
  \chead{%
    \textit{\footnotesize XIV International Conference on Hadron Spectroscopy
      (\weblink[\textit{hadron2011}]{http://www.hadron2011.de}), 13-17 June 2011, Munich, Germany}%
  }
  \rhead{}
  \bigskip
  \begin{center}
    {#3} {#4}\ifthenelse{\equal{#6}{}}{}{\footnote{\weblink[#6]{mailto:#6}}}
    \ifthenelse{\equal{#7}{}}{}{#7} \\
    \textit{#5}
  \end{center}
  \bigskip
}

\renewcommand{\abstract}[1]{%
  \begin{center}
    \begin{minipage}{0.85\textwidth}
      \begin{footnotesize}
        #1
      \end{footnotesize}
    \end{minipage}
  \end{center}
  \bigskip
}

%% file: contribution.tex
%
%
%
%
%
{  


%

\makeatletter
\@ifundefined{c@affiliation}%
{\newcounter{affiliation}}{}%
\makeatother
\newcommand{\affiliation}[2][]{\setcounter{affiliation}{#2}%
  \ensuremath{{^{\alph{affiliation}}}\text{#1}}}
%

\contribution[]
{Exploration of resonance properties in chiral perturbation theory with explicit $U_A(1)$ anomaly}
{Zhi-Hui}{Guo}  
{\affiliation[Department of Physics, Hebei Normal University, 050016 Shijiazhuang, China]{1} \\
 \affiliation[Departamento de F\'{\i}sica, Universidad de Murcia, E-30071 Murcia, Spain]{2} }
{Speaker}
{\!\!$^,\affiliation{1} ^,\affiliation{2}$  and  J.~A.~Oller\affiliation{2} }

%
%

\abstract{%
 We study the resonance properties within chiral perturbation theory by
 explicitly taking into account the $U_A(1)$ anomaly effect. This assures we have
 the appropriate degrees of freedom of low energy QCD in the large $N_C$ limit.
 We calculate the various resonance properties, such as mass, width and residues,
 for the physical case, i.e. $N_C=3$. Then we extrapolate the values of $N_C$ to
 study the trajectories of resonance poles.
}
%

\section{Introduction}

The combination of chiral perturbation theory ($\chi$PT) and non-perturbative methods inspired from
$S$- matrix theory, play an important role in the study of resonances nowadays~\cite{uchpt1}.
A straightforward application of this approach is to determine the properties of the resonance,
such as the  mass and the width. A deeper understanding of the resonance structure can be obtained
by combining the $1/N_C$ expansion of QCD in $\chi$PT~\cite{uchpt2}.
A novel ingredient in our current study is the singlet $\eta_1$, which is massive
even in the chiral limit due to the $U_A(1)$ anomaly effect. This ingredient has been commonly ignored
in the previous works on the study of $N_C$ trajectories of resonance poles~\cite{uchpt2}.
Since it turns out to be the ninth pseudo Goldstone in the chiral
and large $N_C$ limit and becomes a relevant degree of freedom in the low energy QCD for large $N_C$,
it is necessary to take this effect into account to study resonance properties at large $N_C$.
In our study, this gap is filled by using the $U(3)$ $\chi$PT, which incorporates the massive singlet $\eta_1$
as its explicit degree of freedom.

\section{Theoretical setup}

The current study of resonance properties is based on the complete one-loop
calculation of meson-meson scattering within $U(3)$ $\chi$PT by explicitly
including the tree level exchanges of scalar and vector resonances.
The perturbative calculation incorporates not only  the genuine meson-meson scattering diagrams, including the
loops and the resonance exchanges, but also the contributions from the wave function renormalizations, mass renormalizations,
pseudo-Goldstone weak decay constants and $\eta-\eta'$ mixing.
The master formula to unitarize the perturbative $U(3)$ $\chi$PT results is from
a simplified version of N/D method derived in \cite{oller99prd}
\begin{eqnarray}\label{defnd}
{T_J^{I}}(s)^{-1} = N_J^I(s)^{-1}+ g(s)~,
\end{eqnarray}
where $T_J^{I}(s)$ denotes the unitarized partial wave amplitudes with well defined isospin $I$
and angular momentum $J$. $N_J^{I}(s)$ collects the crossed channel
cuts and can be constructed from the perturbative results
\begin{eqnarray}\label{defnfunction}
N_J^I(s) = {T_J^I}(s)^{\rm (2)+Res+Loop}+ T_J^I(s)^{(2)}  \,\, g(s) \, \, T_J^I(s)^{(2)} \,,
\end{eqnarray}
where $T_J^I(s)^{\rm (2)+ Res + Loop}$ stands for the partial wave amplitude from the
perturbative calculation, with the superscripts ${\rm(2),~ Res}$ and Loop
denoting the leading order amplitudes, resonance exchanges and
loop contributions, respectively. The explicit expressions from the perturbative calculation can be
found in Ref.~\cite{guo11prd}. $g(s)$ contains the right hand cuts and its explicit
expression can be also found in \cite{guo11prd} and references therein.

\section{Results and Discussions}
From the unitarized meson-meson scattering amplitudes, one can construct the phase shift,
modulus of the $S$-matrix and invariant mass distribution~\cite{guo11prd}.
Then we fit the various quantities
to the experimental data to get the unknown parameters in our model.
The resonance poles are found on the unphysical Riemann sheets, which can be obtained from the extrapolation
of $T$-matrix on the physical sheet (the first sheet) to the unphyscial ones.
In our study, the pole positions of three kinds of vector resonances are obtained: $\rho(770)$ with $IJ=1\,1$,
$K^*(892)$ with $IJ=\frac{1}{2}\,1$ and $\phi(1020)$ with $IJ=0\,1$.
For the scalar resonances, we find $\sigma$, $f_0(980)$ and $f_0(1370)$ for the $IJ=0\,0$ case,
$a_0(980)$ and $a_0(1450)$ for $IJ=1\,0$ case, $\kappa$ and $K^*(1430)$ for $IJ=\frac{1}{2}\,0$.
The masses and widths of the above mentioned resonances are consistent with the PDG values~\cite{pdg}.
We have also calculated the couplings between the resonances and the pseudo-scalar mesons.
For the detailed result, see Ref.~\cite{guo11prd}.

Then we extrapolate the values of $N_C$ from 3 to larger numbers to study the corresponding
behaviors of resonance poles. In this way, one can learn whether the resonance is a standard $\bar{q}q$
resonance by plotting the $N_C$ trajectories of its pole positions. The basic criteria is that in the framework of
large $N_C$ QCD the mass of a standard $\bar{q}q$ resonance is a constant, while its decay width decreases as $1/N_C$
when $N_C$ approaches to infinity.
For $\rho(770)$, $K^*(892)$, $f_0(980)$, $f_0(1370)$, $a_0(1450)$ and $K^*(1430)$,
we plot the quantities of $\Gamma*N_C$ ($\Gamma$ denotes the width) and mass of the resonances as functions
of $N_C$ in Figs.~\ref{fig:ncwidth} and \ref{fig:resmass} respectively. It is clear that the resonances
appearing in those two figures all behave as the standard $\bar{q}q$ resonances, since their
masses are more or less stable when varying $N_C$ and their decay widths decrease as $1/N_C$ for large values
of $N_C$. However for the very broad resonances $\sigma$ and $\kappa$, we find their pole positions in the
complex $s$ plane approach to the real and negative axis when increasing the values of $N_C$.
In this case, it is not appropriate to interpret the imaginary part of $\sqrt{s}$ as the decay width.
In Fig.~\ref{fig:sigkappa}, we plot the real and imaginary parts of $s_\sigma$ and $s_\kappa$ as functions
of $N_C$, from where one can conclude that $\sigma$ and $\kappa$ resonances in our study do not correspond to the
standard $\bar{q}q$ resonances for large $N_C$.
For $a_0(980)$, both its width and mass increase when increasing $N_C$, indicating
it does not seem to correspond to a standard $\bar{q}q$ resonance in our current study.

\begin{figure}[Htb]
  \begin{center}
    \includegraphics[width=0.6\textwidth]{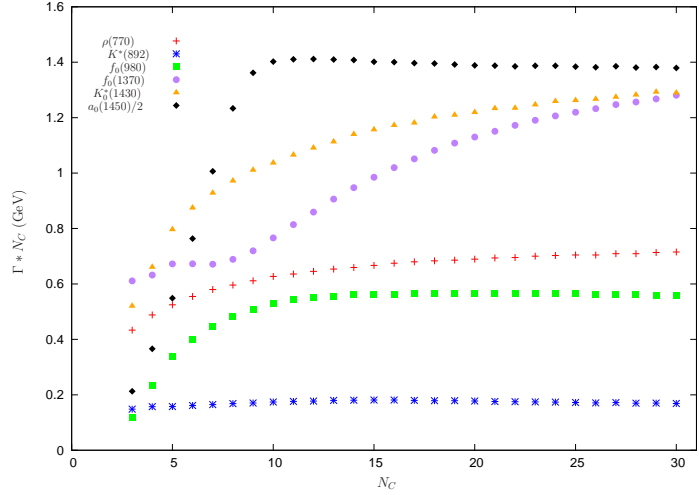}
    \caption{Decay widths of resonances as functions of $N_C$. For $a_0(1450)$, we plot
    the quantity of $\Gamma*N_C/2$. For the others, we plot the quantity of $\Gamma*N_C$ as functions of $N_C$. }
    \label{fig:ncwidth}
  \end{center}
\end{figure}
%
\begin{figure}[Htb]
  \begin{center}
    \includegraphics[width=0.6\textwidth]{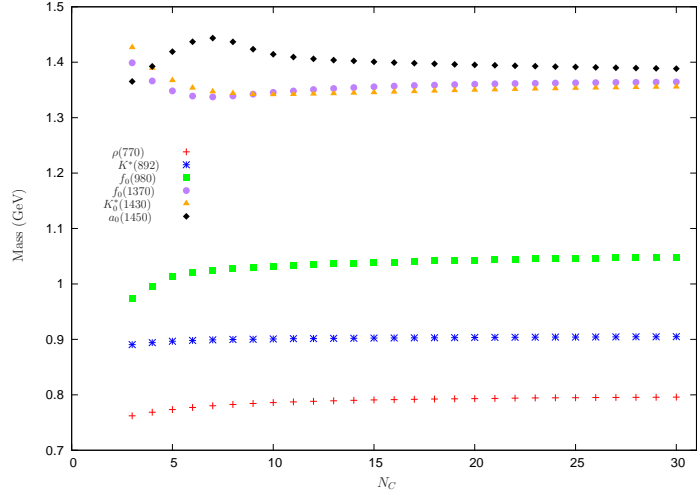}
    \caption{Masses of resonances as functions of $N_C$.}
    \label{fig:resmass}
  \end{center}
\end{figure}
%
\begin{figure}[htb]
  \begin{center}
    \includegraphics[width=0.8\textwidth]{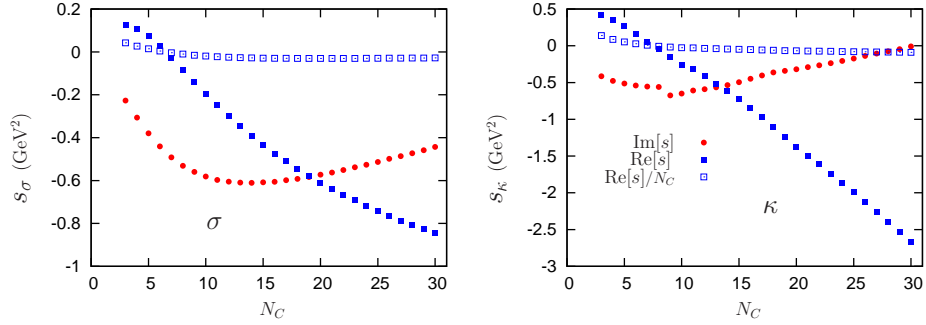}
    \caption{$N_C$ trajectories for the imaginary and real parts of $\sigma$ pole $s_\sigma$ and $\kappa$ pole $s_\kappa$. }
    \label{fig:sigkappa}
  \end{center}
\end{figure}
%

\acknowledgements{%
This work is partially funded by the DGICYT grant FPA2010-17806,
the Fundaci\'on S\'eneca grant with Ref.~11871/PI/09,
the EU-Research Infrastructure Integrating Activity ``Study of Strongly Interacting Matter" (HadronPhysics2, grant No.227431)
under the Seventh Framework Programme of EU, the Consolider-Ingenio 2010 Programme CPAN (CSD2007-00042),
Natural Science Foundation of Hebei Province with contract No. A2011205093
and Doctor Foundation of Hebei Normal University with contract No. L2010B04.
}


%

}  


%% file: hadron2011.bbl
\begin{thebibliography}{99}


\bibitem{uchpt1} J.~Oller and E.~Oset, Nucl.~Phys.~{\bf A620} (1997) 438; (E)$-ibid$  A {\bf 652}(1999) 407;
M.~Albaladejo and J.~A.~Oller, Phys.~Rev.~Lett.~{\bf101} (2008) 252002;
I.~Caprini, G.~Colangelo and H.~Leutwyler, Phys.~Rev.~Lett.~{\bf 96} (2006) 132001.

\bibitem{uchpt2}
J.~R.~Pel\'aez, Mod.~Phys.~Lett.~{\bf A19} (2004) 2879;
Z.~H.~Guo, J.~J.~Sanz-Cillero and H.-Q.~Zheng, JHEP {\bf 06} (2007) 030;
J.~Nieves, A.~Pich and E.~Ruiz Arriola, arXiv: 1107.3247 [hep-ph];
L.~Y.~Dai, X.~G.~Wang and H.~Q.~Zheng, arXiv: 1108.1451 [hep-ph].

\bibitem{oller99prd} J.~A.~Oller and E.~Oset, Phys.~Rev.~{\bf D60} (1999) 074023;
J.~A.~Oller, Phys. Lett. B {\bf 477} (2000) 187;
J.~A.~Oller and U.~G.~Meissner, Phys.\ Lett.\  B {\bf 500} (2001) 263.

\bibitem{guo11prd}
  Zhi-Hui~Guo and J.~A.~Oller,  Phys.~Rev.~{\bf D84} (2011) 034005;
  Z.~H.~Guo, J.~Prades and J.~A.~Oller, Nucl.\ Phys.\ Proc.\ Suppl.\ B {\bf 207-208} (2010) 184.

\bibitem{pdg} K. Nakamura, {\it et al.}, (Particle Data Group), J. Phys. G 37 (2010) 075021.


\end{thebibliography}
